\documentstyle[12pt]{article}
\catcode`@=11

\def\beqra{\begin{eqnarray}} \def\eeqra{\end{eqnarray}}
\def\beqast{\begin{eqnarray*}} \def\eeqast{\end{eqnarray*}}
\def\beq{\begin{equation}}      \def\eeq{\end{equation}}
\def\be{\begin{enumerate}}   \def\ee{\end{enumerate}}

\def\fo{\hbox{{1}\kern-.25em\hbox{l}}}

\def\om{\omega}

\def\rta{\rightarrow}

\def\nd{\noindent}
\def\ch{\@startsection{section}{1}{\z@}{-3ex plus-1ex minus-.2ex}%
        {2ex plus.2ex}{\large\sc}}

\def\raisenot{\raise .5mm\hbox{/}}
\def\nota{\ \hbox{{$a$}\kern-.49em\hbox{/}}}
\def\notA{\hbox{{$A$}\kern-.54em\hbox{\raisenot}}}
\def\notb{\ \hbox{{$b$}\kern-.47em\hbox{/}}}
\def\notB{\ \hbox{{$B$}\kern-.60em\hbox{\raisenot}}}
\def\notc{\ \hbox{{$c$}\kern-.45em\hbox{/}}}
\def\notd{\ \hbox{{$d$}\kern-.53em\hbox{/}}}
\def\notbd{\ \hbox{{$D$}\kern-.61em\hbox{\raisenot}}} 
\def\note{\ \hbox{{$e$}\kern-.47em\hbox{/}}}
\def\notk{\ \hbox{{$k$}\kern-.51em\hbox{/}}}
\def\notp{\ \hbox{{$p$}\kern-.43em\hbox{/}}}
\def\notq{\ \hbox{{$q$}\kern-.47em\hbox{/}}}
\def\notW{\ \hbox{{$W$}\kern-.75em\hbox{\raisenot}}}
\def\notz{\ \hbox{{$Z$}\kern-.61em\hbox{\raisenot}}}

\def\notpa{\hbox{{$\partial$}\kern-.54em\hbox{\raisenot}}}

\def\rf#1{$^{#1}$}

\def\7#1#2{\mathop{\null#2}\limits^{#1}}        
\def\5#1#2{\mathop{\null#2}\limits_{#1}}        

\def\inbar{\vrule height1.5ex width.4pt depth0pt}
\def\IB{\relax{\rm I\kern-.18em B}}
\def\IC{\relax\leavevmode\hbox{\,$\inbar\kern-.3em{\rm C}$}}
\def\ID{\relax{\rm I\kern-.18em D}}
\def\IE{\relax{\rm I\kern-.18em E}}
\def\IF{\relax{\rm I\kern-.18em F}}
\def\IG{\relax\leavevmode\hbox{\,$\inbar\kern-.3em{\rm G}$}}
\def\IH{\relax{\rm I\kern-.18em H}}
\def\II{\relax{\rm I\kern-.18em I}}
\def\IK{\relax{\rm I\kern-.18em K}}
\def\IL{\relax{\rm I\kern-.18em L}}
\def\IM{\relax{\rm I\kern-.18em M}}
\def\IN{\relax{\rm I\kern-.18em N}}
\def\IO{\relax\leavevmode\hbox{\,$\inbar\kern-.3em{\rm O}$}}
\def\IP{\relax{\rm I\kern-.18em P}}
\def\IQ{\relax\leavevmode\hbox{\,$\inbar\kern-.3em{\rm Q}$}}
\def\IR{\relax{\rm I\kern-.18em R}}
\def\sed{\hbox{{\sf S}\kern-.4em\hbox{\sf S}}}
\def\ZZ{\relax{\sf Z\kern-.4em Z}}
\def\smIR{\hbox{{\footnotesize\rm I}\kern-.2em\hbox{\footnotesize\rm R}}}
\def\smIO{\ \hbox{{\footnotesize\rm I}\kern-.4em\hbox{\footnotesize\bf O}}}
\def\smIQ{\ \hbox{{\footnotesize\rm I}\kern-.5em\hbox{\footnotesize\bf Q}}}
\def\IGa{\relax{\rm I}\kern-.18em\Gamma}
\def\IPi{\relax{\rm I}\kern-.18em\Pi}
\def\IQt{\relax\leavevmode\hbox{$\kern.3em\inbar\kern-.3em\Theta$}}
\def\IOm{\relax\hbox{$\kern3.48pt\inbar\kern1.8pt\inbar\kern-5.28pt\Omega$}}

\def\ca#1{\relax\ifmmode {\cal#1} \else$\cal#1$\fi}     
\def\Sf#1{\relax\ifmmode\hbox{\sf#1}\else{\sf#1}\fi}    
\def\fibby{\ifcase\@ptsize                      
                \font\tenrm=cmfib8\or           
                \font\elvrm=cmfib8 scaled\magstephalf\or        
                \font\twlrm=cmfib8 scaled\magstep1 \fi}         
\def\TeXey{\ifcase\@ptsize\or\or                
                \font\twlrm=cmr10 scaled\magstep1       
                \font\twlmi=cmmi10 scaled\magstep1      
                \font\twlit=cmti10 scaled\magstep1      
                \font\twlbf=cmbx10 scaled\magstep1\fi}  

\def\ch{\@startsection{section}{1}{\z@}{-3ex plus-1ex minus-.2ex}%
        {2ex plus.2ex}{\large\sc}}
\def\sch{\@startsection{subsection}{2}{\z@}{-1.5ex plus-1ex minus-.2ex}%
        {1pt plus.2ex}{\sc}}
\def\ssch{\@startsection{subsubsection}{3}{\z@}{-1ex plus-1ex minus-.2ex}%
        {1pt plus.2ex}{\small\sc}}
\def\seceq{\@addtoreset{equation}{section}
        \def\theequation{\thesection.\arabic{equation}}}        

\def\con{\ifmmode \hbox{\bf*} \else{\bf*}\fi}   
\def\scon{\ifmmode \hbox{\footnotesize\rm\bf*} \else{\footnotesize\rm\bf*}\fi}

\def\0#1{\relax\ifmmode\mathaccent"7017{#1}
        \else\accent23#1\relax\fi}              

\def\haf{\frac{1}{2}}

\def\place#1#2#3{\vbox to0pt{\kern-\parskip\kern-7pt
                             \kern-#2truein\hbox{\kern#1truein #3}
                             \vss}\nointerlineskip}

\def\illustration #1 by #2 (#3){\vbox to #2{\hrule width #1 height 0pt depth
0pt
                                       \vfill\special{illustration #3}}}

\def\scaledillustration #1 by #2 (#3 scaled #4){{\dimen0=#1 \dimen1=#2
           \divide\dimen0 by 1000 \multiply\dimen0 by #4
            \divide\dimen1 by 1000 \multiply\dimen1 by #4
            \illustration \dimen0 by \dimen1 (#3 scaled #4)}}

\catcode`@=12

\begin{document}

\hfill{DOE-ER-40757-030}

\hfill{CPP-93-30}

\hfill SMU-HEP-94/02

\hfill hep-ph/9402262

\vspace{12pt}
\begin{center}
 \large{\bf STANDARD MODEL DECAYS OF TAU INTO THREE CHARGED LEPTONS}\\

\normalsize

\vspace{24pt}

Duane A. Dicus

\vspace{12pt}
{\it Center for Particle Physics and Department of Physics\\
University of Texas, Austin, Texas 78712}\\

\vspace{24pt}
Roberto Vega

\vspace{12pt}
{\it Physics Department  \\
Southern Methodist University,  Dallas, Texas 75275}

\vspace{36pt}
ABSTRACT
\end{center}

Analytic expressions are given for the $\tau$ decays into three charged
leptons,
$\tau\rta \ell\ell\overline \ell\nu_\tau\overline \nu_\ell$,
where the $\ell$ are combinations of electrons and muons, and for the
radiative decays $\tau\rta \gamma\ell\nu_\tau\overline \nu_\ell$.
The branching ratios are sensitive functions of whether the final state
$\ell$ are muons or electrons.

\def\im{{\rm Im}}

\pagebreak
\baselineskip=20pt
\setcounter{page}{1}

The branching ratio for the process
$\mu^+\rta e^+e^+e^- \nu_e\overline \nu_\mu$
was first measured more than thirty years ago\rf{1-4}.
The measurements are in good agreement
with the standard model theoretical predictions\rf5.
More recently, the primary
interest in this process is as a background for processes
that would signal new physics beyond the standard model such as
$\mu^+\rta e^+e^+e^-$.

The study of $\tau$ decays is rapidly reaching the point where similar studies
could be done.  At CLEO, for example, enough $\tau$s' may have already
been produced to allow measurement of the branching ratios for the four
leptonic final states,
$e^+e^+e^-\nu_e        \overline \nu_\tau~~,
{}~~\mu^+e^+e^-\nu_\mu   \overline \nu_\tau~~,
{}~~e^+\mu^+\mu^-\nu_e  \overline \nu_\tau$ and
$\mu^+\mu^+\mu^-\nu_\mu\overline \nu_\tau$;
or to place limits on the branching ratios for the corresponding
lepton number violating processes which do not involve neutrinos
in the final states\rf6.
The purpose of this note is to give the standard model
branching ratios for
$\ell^+\ell^+\ell^-\nu_\ell\overline \nu_\tau$, where $\ell=e$ or $\mu$; and
to present an analytic expression for the square of the matrix element
which can be easily incorporated into Monte Carlo simulations.

The processes proceed through the Feynman diagrams given in Figure 1 where the
last two diagrams are needed only for the cases of identical particles in the
final state.  Each term in the square of the matrix element is of the form
\beqast
Tr\,\gamma^\beta\left(1-\gamma_5\right)\gamma\cdot p_5\,\gamma^\alpha
\left(1-\gamma_5\right)A\times\;Tr\;\gamma_\alpha\left(1-\gamma_5\right)
\gamma\cdot p_4\,\gamma_\beta\left(1-\gamma_5\right)B
\eeqast
where $A$ and $B$ are strings of gamma matrices and $p_4$ and $p_5$
are the momenta of the neutrinos. If these traces are done directly
the result involves many terms.  However, the result can be made shorter
by first introducing an identity matrix on either side of $A$ and of
$B$, and Fierz rearranging both.  The expression then becomes
\beqast
Tr\,A\gamma_\sigma\left(1+\gamma_5\right)\;Tr\,\gamma_\beta
\left(1-\gamma_5\right)\gamma\cdot p_5\,\gamma_\alpha\left(1-\gamma_5\right)
\gamma^\sigma   \\
\times\,Tr\,B\gamma_\rho\left(1+\gamma_5\right)\;Tr\,\gamma^\alpha
\left(1-\gamma_5\right)\gamma\cdot p_4\,\gamma^\beta\left(1-\gamma_5\right)
\gamma^\rho   \\
\sim\;p_5^\rho\,p_4^\sigma\;Tr\,A\gamma_\sigma\left(1+\gamma_5\right)\;
Tr\,B\gamma_\rho\left(1+\gamma_5\right)
\eeqast
and the squared matrix element is greatly simplified.  Further, since
this is the only dependence on $p_4$ and $p_5$, the phase space integrals
of the neutrinos can be easily done.

  The total rate for the process $\tau^+(P)
\rta\mu^+(p_1)+e^+(p_2)+  e^-(p_3)+\overline{\nu}_\tau+\nu_\mu$
is given by

\beq
\Gamma = \left[\frac{G}{\sqrt{2}}\right]^2\, \frac{\propto^2}{(2\pi)^8}\,
\frac{128}{3}\, \frac{1}{M}
 \int\,\frac{d^3p_1}{2E_1}\,\frac{d^3p_2}{2E_2}\,\frac{d^3p_3}{2E_3}\,
\Theta(Q^2)\Theta(Q^0)\,\frac{1}{s^2}\, T
\eeq
where $Q^\mu = P^\mu-p_1^\mu-p_2^\mu-p_3^\mu$,
$P^2=M^2,\; p_1^2=m_\mu^2,~~p_2^2=m_e^2=p_3^2$,
and $s=(p_2+p_3)^2$.  The $\Theta$
functions are what remain after analytically computing the integrals
over the neutrino momenta.
The rest of the squared matrix element, $T$, is given in Appendix
I.  Expressions for the rate to the other final states can be
obtained by interchanging $m_\mu$ and $m_e$ or by setting them
equal.  In addition, for the identical lepton cases, the expression for
$\frac{1}{s^2}T$ has to be generalized as explained in the appendix.
The rate for $\mu^+\rta e^+e^+e^-\nu_e\overline \nu_\mu$
can also be obtained from (1) by
setting $m_\mu$ to $m_e$ and then $M$ equal to $m_\mu$.

The branching ratios are given by the following table

\begin{center}
\begin{tabular}{ll}

$ \mu^+\rta e^+e^+e^-\nu_e\overline\nu_\mu  \qquad$ & $3.60\pm0.02\times 10^{-5
}$ \\[4pt]
$\tau^+\rta e^+e^+e^-\nu_e\overline\nu_\tau \qquad$   & $4.15\pm0.06\times
10^{-5}$\\[4pt]
$\tau^+\rta \mu^+e^+e^-\nu_\mu\overline\nu_\tau \qquad$ & $1.97\pm 0.02\times
10^{-5}$
\\[4pt]
$ \tau^+\rta e^+\mu^+\mu^-\nu_e\overline\nu_\tau\qquad$ & $1.257 \pm
0.003\times
10^{-7}$\\[4pt]
$\tau^+\rta  \mu^+\mu^+\mu^-\nu_\mu\overline\nu_\tau\qquad$ & $1.190\pm 0.002
\times
10^{-7}$\\[4pt]

\end{tabular}
\end{center}
where the errors are due to inaccuracies in the numerical integration.
These numbers assume the branching ratio for $\tau\rta e\nu_\tau\nu_e$
is 0.177.

For the tau cases where the pair produced are electrons, the branching ratio is
of the same order as the muon decay.  If the pair produced are muons, however,
the branching ratio is two orders of magnitude smaller.  This seems to be just
the effect of the $1/s^2$ in Eq. (1); for electrons $s$ can be a lot smaller
than for muons.  $(s\geq 4m^2$ where $m$ is the electron or muon mass.)  This
means that any experimental cut on the minimum value of invariant mass of the
produced pair will drastically reduce the number of events.  This also means
that the cross terms between the  first and second diagrams of Fig. 1 and the
third and fourth diagrams make only a small contribution to the rate because
these cross terms go as $1/ss'$, where $s'=(p_1+p_3)^2$, and $s$ and $s'$
cannot be small simultaneously.  These observations were verified by
our numerical results.

An analysis of the angular distributions of the final state
leptons reveals strong correlations.  For the case with three
electrons in the final state we have plotted, in
Fig. 2a, the distributions in  the angles between the two
like-sign and the opposite-sign electrons.  In both cases
the distributions peak when the two momenta are parallel.  Similar
results are obtained for the final state
$\mu^+e^+e^-\nu_\mu   \overline \nu_\tau~~$, but the correlations
disappear for the final states
$e^+\mu^+\mu^-\nu_e  \overline \nu_\tau$ and
$\mu^+\mu^+\mu^-\nu_\mu\overline \nu_\tau$.  Again these results
are simply a consequence of the $1/s^2$ term dominating the
dynamics.   In Fig. 2b we have plotted the distributions
in the total energy $E_c$ of the final state charged leptons.  For
comparison we also display the same distribution for just the phase
space without any dynamics.  The dynamics shift the $E_c$ distribution
to lower values.  The shift to lower values increases with decreasing
mass of the final state leptons.

For completeness we also include an expression for the radiative decay
$\tau(P)\rta \ell(p_2)+\gamma
(k)+\nu_\tau+\nu_\ell$ for $\ell=e$ or $\mu$.  With the neutrinos again
integrated
out the rate is given by

\beq
\Gamma = \frac{G^2\propto}{(2\pi)^2}\;\frac{4}{3}\;\frac{1}{M} \int
\,\frac{d^3k}{2\om}\;\frac{d^3p_2}{2E_2}\, \Theta \, (K^2)\Theta(K^0)T_\gamma
\eeq
where $K^\mu=P^\mu-k^\mu-p_2^\mu$ and $T_\gamma$ is given in Appendix II.  The
photon must be visible and thus the photon energy has some nonzero lower limit
$\om_0$.  This form would seem to be useful if further experimental cuts need
to be applied to the photon or charged lepton.  If, however, no further cuts
are required, then the integrals can be done analytically and the branching
ratio to $\tau\rta \ell +\nu_\tau+\nu_\ell$ is given by\rf7

\beqra
R &=& \frac{\alpha}{3\pi}\,
\left\{\left(\ell n\,\frac{M}{m} - \frac{17}{12}\right)\left[
6\,\ell n\,\frac{1}{y_0}-6(1-y_0)-(1-y_0)^4\right]\right. \nonumber \\[5pt]
&&+\, 3\left[L(1)-L(y_0)\right]-\haf\left[6+(1-y_0)^3\right](1-y_0)\ell
n(1-y_0)
\nonumber \\[5pt]
&& +\,\left. \frac{1}{48}\,(1-y_0)(125+45y_0-33y_0^2+7y_0^3)\right\},
\eeqra
where $y_0=\omega_0/\omega_{{\rm Max}} $ and $L$ is a Spence function defined
with an overall sign such that $L(1)=-\pi^2/6$. A graph of the branching ratio
given by this equation is shown in Fig. 3 for the three processes $\mu\rta
e+\gamma+\nu+\overline \nu$, $\tau\rta e+\gamma+\nu+\overline \nu$, and
$\tau\rta
\mu+\gamma+ \nu+\overline \nu$.

In conclusion we have found that with about one million $\tau$s'
the standard model branching
ratio for $\tau$ decays to the final states
$e^+e^+e^-\nu_e        \overline \nu_\tau$ and
$\mu^+e^+e^-\nu_\mu   \overline \nu_\tau$ may be measurable.
The distributions in the angles between the final state leptons
are strongly peaked at zero and stringent angular separation
cuts can drastically reduce the rates.

The authors would like to thank Ryszard Stroynowski and Igor Volobouev
for stimulating this work and for useful discussions.  R.V.
would like to thank the Lightner-Sams Foundation for support.
This research was supported in part by DOE grants DE-FG05-92ER-40722
and DE-FG03-93ER40757.

\pagebreak
\nd
{\bf APPENDIX I}

For the decay $\tau^+(P)\rta\mu^+(p_1)+e^+(p_2)+ e^-(p_3) +\overline
\nu_\tau+\nu_\mu$
define

\begin{center}
\begin{tabular}{lcl}

$Q^\mu$ &$=$& $P^\mu-p_1^\mu-p^\mu_2-p_3^\mu$ \\[3pt]
$s$&$=$& $(p_2+p_3)^2$ \\[3pt]
$s'$ &$=$& $(p_1+p_3)^2 $\\[3pt]
$t$ &$=$ & $p_1\cdot p_2$ \\[3pt]
$u$ &$=$ & $Q\cdot P$ \\[3pt]
$s_1$ &$=$ & $P\cdot p_1$ \\[3pt]
$s_2$ &$=$ & $P\cdot p_2$ \\[3pt]
$s_3$ &$=$ & $P\cdot p_3$ \\[3pt]
$u_1$ &$=$ & $Q\cdot p_1$ \\[3pt]
$u_2$ &$=$ & $Q\cdot p_2$ \\[3pt]
$u_3$ &$=$ & $Q\cdot p_3$ \\[3pt]

\end{tabular}
 \end{center}

\nd
Obviously there are relations among these invariants but the following analytic
expression is
simpler if we use an overcomplete set.  In terms of these the expression for
$T$ in Eq. (1) is
given by
\beqast
T &=& \frac{1}{s-2(s_2+s_3)}\Big[ -(Q^2s_1+2uu_1)(m_e^2+s_2+s_3) \\[3pt]
& + &\haf\, Q^2\left(2s_2t+s_3\left(s'-m_e^2-m_\mu^2\right)\right) +
2u_1(u_2s_2
+u_3s_3)\Big] \\[3pt]
&-& \frac{2}{[s-2(s_2+s_3)]^2}\left[\haf
Q^2(2s_1-2t-s'+m_e^2+m_\mu^2)+2u_1(u-u_2-u_3)
\right] \\[3pt]
&& \times~ \left[s^2_5+s^2_4+\haf sM^2+m_e^2(s_2+s_3)\right] \\[3pt]
&+& \frac{1}{s+s'+2t-m_e^2-m_\mu^2}\,\left[(Q^2s_1+2uu_1) \left(t+
\frac{s'}{2}-m_e^2-\frac{m^2_\mu}{2}\right)\right. +t(Q^2s_2+2uu_2) \\[3pt]
&&~~~~+(Q^2s_3+2uu_3)\left.\haf(s'-m^2_e-m^2_\mu)\right] \\[3pt]
&+& \frac{2}{\left[s+s'+2t
-m_e^2-m_\mu^2\right]^2}\left[Q^2(s_1+s_2+s_3)+2u(u_1+u_2+
u_3)\right] \\[3pt]
&& \times~\left[-t^2-\frac{1}{4}(s'-m^2_e-m_\mu^2)^2-\haf\,sm_\mu^2+\haf
m_e^2(2t+s'-m_e^2
- m_\mu^2) \right] \\[3pt]
&+& \frac{1}{s-2(s_2+s_3)}\,\frac{1}{s+s'+2t-m_e^2-m_\mu^2}\\[3pt]
&\times& \bigg[-s(s_2+s_3)(Q^2m^2_\mu+2u^2_2) \\[3pt]
&+& \left(Q^2t+2u_1u_2\right)\left(ss_1+s_3 \left(s'-2
t-s+m_e^2-m_\mu^2\right)+2m_e^2s_2\right)\\[3pt]
&+& \haf\,\left(Q^2\left(s'-m_e^2-m^2_\mu\right)+4u_1u_3\right) \\[3pt]
&& \times~\left(ss_1+s_2\left(2t-s'-s+m_\mu^2+3 m_e^2\right)+2m_e^2s_3\right)
\\[3pt]
&+& 2\left(Q^2s_1+2uu_1\right)\left(2ts_3+s_2\left(s'-m_e^2-m_\mu^2\right)-s
\left(s_1+
m_e^2\right)\right) \\[3pt]
&+& 2\left(Q^2  m_e^2+2u^2_4\right)\left(s_3\left(s'-m_e^2-m_\mu^2\right)
-m_e^2s_1\right) \\[3pt]
&+& 2\left(Q^2 m_e^2+2u^2_5\right)\left(2ts_2-m_e^2s_1\right) \\[3pt]
&+& \left(Q^2 \frac{s}{2}-Q^2m_e^2+2u_2u_3\right)
\left(2ss_1-4m_e^2s_1-4ts_3-2s_2\left(s'-m_e^2-m_\mu^2\right)\right) \\[3pt]
&+& \left(Q^2s_2+2uu_2\right)\left(\left(s_2-s_3\right)
\left(s'-m_e^2-m_\mu^2\right)
+2m_e^2t-ss_1+\haf\left(2m_e^2-s\right)\left(s'-m_e^2-m_\mu^2\right)\right)
\\[3pt]
&+& \left(Q^2s_3+2uu_3\right)\left(2t \left(s_3-s_2+ 2m_e^2 \right)-s
\left(s_1+t\right)+
m_e^2\left( s'-m_e^2-m_\mu^2\right)\right) \\[3pt]
&+& \haf\,s\left(Q^2M^2+2u^2\right) \left(2t+s'-m_e^2-m_\mu^2\right) \\[3pt]
&+& \left.6Q^2\left(-m_e^2\left(s_2+s_3\right)\left(2t+s'-m_e^2-
m_\mu^2\right)+ss_3t+2m_e^2ss_1+\haf
ss_2\left(s'-m_e^2-m_\mu^2\right)\right)\right]\,.
\eeqast
If all final leptons are the same flavor set $m^2_\mu=m^2_e$ and replace
$\displaystyle{\frac{1}{s^2}T}$ by
$$
\haf\,\frac{1}{s^2}\, T+\haf\,\frac{1}{s^{'2}}\, T\left(s_1\leftrightarrow
s_2,u_1\leftrightarrow u_2, s\leftrightarrow s'\right) +\haf\,\,\frac{1}{s}
\,\frac{1}{s'}\, T'
$$
where $T'$ is given by
\beqast
&-& \frac{1}{s-2(s_2+s_3)}\;\frac{1}{s'-2(s_1+s_3)} \\[3pt]
&\times & \bigg[ \left(Q^2t+2u_1u_2\right) \left(4s_3^2-2m_e^2M^2\right)
\\[3pt]
&+& \left(Q^2\left(\frac{s'}{2}-m_e^2\right)+2u_1u_3\right)\left(-2s_2s_3+s
\left(s_3+M^2\right) -m_e^2\left(2M^2+s_2+s_3\right)\right)\\[3pt]
&+&\left(Q^2s_1+2uu_1\right)\left(2s_2s_3-ss_3-\haf
M^2s+2m_e^2\left(s_2+s_3\right)\right)
\\[3pt]
&+&\left(Q^2\left(\frac{s}{2}-m_e^2\right)+2u_2u_3\right)\left(-2s_1s_3+s's_3+s'M^2-m_e^2
\left(2M^2+s_1+s_3\right)\right)\\[3pt]
&+& \left(Q^2s_2+2uu_2\right)\left(2s_1s_3-s's_3-\haf\, M^2s'+2m_e^2
\left(s_1+s_3\right)\right) \\[3pt]
&+&\left(Q^2m_e^2+2u_3^2\right)\left(-2t\left(s_3+M^2\right)-m_e^2
\left(s_1+s_2
\right)\right) \\[3pt]
&+&
\left(Q^2s_3+2uu_3\right)\left(t\left(M^2+4s_3\right)+m_e^2\left(2s_2+2s_1+t+\haf
s+\haf s'-M^2-m_e^2\right)\right)\\[3pt]
&+&
\left(Q^2M^2+2u^2\right)\left(-2ts_3+m_e^2\left(2s_3-2t-s-s'+2m_e^2\right)\right)
\\[3pt]
&+& 6Q^2\left(-2ts^2_5+m_e^2\left(-2s_2s_3-2s_1s_3-2s_1s_2+\haf
s'\left(s_2+s_3\right)
+\haf s\left(s_1+s_3\right) \right.\right.\\[3pt]
&&
\left.\left.\left.~~~~~~+~M^2\left(2t+s'+s-2m_e^2\right)\right)\right)\right]
\\[3pt]
&-& \frac{1}{[s+s'+2t-2m_e^2]^2 } \\[3pt]
&& \times\bigg[\left(Q^2(s_1+s_2)+2u(u_1+u_2)\right)\left(-2t(s+s')+
2m_e^2\left(s+s'-2m_e^2 \right)\right) \\[3pt]
&& + 4\left(Q^2s_3+2uu_3\right)t\left(t-2m_e^2\right)\bigg] \\[3pt]
&-& \frac{1}{s+s'+2t-2m_e^2}~\frac{1}{s-2(s_2+s_3)}  \\[3pt]
&&  \times\bigg[\left(Q^2
m_e^2+2u^2_2\right)ss_3+\left(Q^2m_e^2+2u^2_4\right)\left(
-s's_3+m_e^2\left(s_1+s_3\right)\right)  \\[3pt]
&+& \left(Q^2 m_e^2+2u^2_5\right)\left(-2ts_2+m_e^2(s_1+s_2)\right) \\[3pt]
&-& \left(Q^2 M^2+2u^2\right)st \\[3pt]
&+& \left(Q^2
t+2u_1u_2\right)\left(s_3\left(s-s'\right)-m_e^2\left(s_2+3s_3\right)\right)
\\[3pt]
&+& \left(Q^2\left(\frac{s'}{2}-m_e^2\right) + 2u_1u_3\right)\left(s's_2-ss_1
+m_e^2(s_2-s_3)\right) \\[3pt]
&+& \left(Q^2 s_1+2uu_1\right)\left(-2ts_3-s's_2
+ss_1+m_e^2(s+4s_3-2s_2)\right)
\\[3pt]
&+& \left(Q^2\left(\frac{s}{2}-m_e^2\right) +
2u_2u_3\right)\left(2ts_3+s's_2-ss_1
+ m_e^2\left(2s_1-s_2-s_3\right)\right) \\[3pt]
&+& \left(Q^2 s_2+2uu_2\right)\left(ss_1-s's_2+m_e^2\left(2s_3-\haf s-\haf
s'-t+m_e^2\right)\right) \\[3pt]
&+& \left(Q^2 s_3+2uu_3\right)\left(t (s+2s_2 )+m_e^2 (2s_3-2s_2+\haf s-\haf
s'-t+m_e^2 )\right) \\[3pt]
&+& 6\,Q^2 \left(-sts_3+m_e^2\left(t\left(s_2+s_3\right) +\haf
s'\left(s_2+s_3\right)
+\haf s\left(s_2-s_3-2s_1\right)\right) \right.\\[3pt]
&& \left.\hspace{1in} -~m_e^2\left(s_2+s_3\right)\Big)\right]\\[4pt]
 &-& \frac{1}{s+s'+2t-2m_e^2}~\frac{1}{s'-2(s_1+s_3)}  \\[3pt]
&\times& \bigg[ \left(Q^2m^2_e+2u^2_2\right)\left(-ss_3+m_e^2(s_2+s_3)\right)+
\left(Q^2m_e^2+ 2u^2_4\right)s's_3 \\[3pt]
&+& \left( Q^2m_e^2+2u^2_5\right)\left(-2ts_1+m_e^2(s_1+s_2)\right)-\left( Q^2
M^2+2u^2\right)s't \\[3pt]
&+& \left(Q^2t + 2u_1u_2\right)\left(s_3(s'-s)-m_e^2(s_1+3s_3)\right) \\[3pt]
&+& \left(Q^2\left(\frac{s'}{2}-m_e^2\right) +
2u_1u_3\right)\left(2ts_3-s's_2+ss_1+m_e^2(2s_2-s_1-s_3)\right) \\[3pt]
&+& \left(Q^2s_1+2uu_1\right)\left(s's_2-ss_1+m_e^2\big(2s_3-\haf s-\haf
s'-t + m_e^2\big)\right)   \\[3pt]
&+& \left(Q^2 \left(\frac{s}{2}-m_e^2\right)+2u_2u_3\right)\left(
ss_1-s's_2+m_e^2\left(s_1-s_3 \right)\right) \\[3pt]
&+& \left(Q^2s_2 + 2uu_2\right)\left(-
2ts_3+s's_2-ss_1+m_e^2\left(s'+4s_3-2s_1\right)\right) \\[3pt]
&+& \left(Q^2
s_3+2uu_3\right)\left(t\left(s'+2s_1\right)+m_e^2\big(-t+2s_3-2s_1+\haf
s'-\haf s+m_e^2\big)\right) \\[3pt]
&+&   6 Q^2 \bigg(-s'ts_3+m_e^2\left(t(s_1+s_3\right)+\haf
s'(s_1-2s_2-s_3)+\haf
s(s_1+s_3)\big)  \\[3pt]
&&\hspace{2in}~~ +~m_e^4\left(-s_1-s_3\right)\bigg)\bigg]
\eeqast

\nd
{\bf APPENDIX II}

The decay $\tau(P)\rightarrow \ell(p_2)+\gamma(k)+\nu+\overline{\nu}$ is given
in the text
in terms of $T_\gamma$.  If the energy of the charged lepton, the energy of the
photon, the
mass of $\tau$, and the mass of $\ell$ are denoted by $E_2,\om,M$, and $m$, and
$k\cdot
p_2$ is named $u$, then $T_\gamma$ is given by
\beqast
T_\gamma &=& 4u^2\left(\frac{1}{M\om}+\frac{1}{\om^2}\right) \\[3pt]
&& + u\left[\frac{1}{\om}\left(-5M+3\frac{m^2}{M}-12E_2\right)
+\frac{1}{\om^2}\left(3 M^2-8ME_2+3m^2\right)-4\right] \\[3pt]
&+& \frac{1}{u}\,\bigg[-4\om^2 M^2+\om\left(3M^3-12M^2E_2-5Mm^2\right)  \\[3pt]
&& +\frac{1}{\om}\left(6M^3E_2^2-4M^2m^2E_2-8M^2E_2^3+6Mm^2E_2^2\right) \\[3pt]
&& + 6M^3E_2+ 3M^2m^2-16M^2E_2^2-2Mm^2E_2+3m^4\bigg] \\[3pt]
&+& \frac{1}{u^2}\bigg[ 4M^2m^2\om^2 +\om\left(-3M^3m^2+8M^2m^2E_2-3Mm^4\right)
\\[3pt]
&& -3M^3m^2E_2+2M^2m^4+4M^2m^2E_2^2-3Mm^4E_2\bigg] \\[3pt]
&+& 4M\om +2M^2+16ME_2+2m^2 \\[3pt]
&+& \frac{1}{\om} \left[-3M^3+2M^2E_2-3Mm^2+16 ME_2^2- 6m^2E_2\right] \\[3pt]
&+& \frac{1}{\om^2} \left[-3M^3E_2+2M^2m^2 + 4M^2E^2_2-3Mm^2E_2\right]
\eeqast

\vspace{10pt}
\centerline{\bf REFERENCES }

\be
\item J. Lee and N.P. Samios, Phys. Rev. Lett. {\bf 3}, 55 (1959).

\item I.I. Gurevich, B.A. Nikol'skii and L.V. Surkova, Sov. Phys. - JETP {\bf
10}, 225
(1960).

\item R.R. Crittenden, W.D. Walker and J. Ballam, Phys. Rev. {121}, 1823
(1961).

\item W. Bertl, et al., Nucl. Phys. {\bf B260}, 1 (1985).

\item D. Yu Bardin, Ts. G. Istatkov and G.B. Mitsel'makher, Sov. J. Nucl. Phys.
{\bf 15},
161 (1972).

\item Ryszard Stroynowski, private communication.

\item T. Kinoshita and A. Sirlin, Phys. Rev. Lett. {\bf 2}, 177 (1959).
\ee

\pagebreak

\centerline{\bf FIGURE CAPTIONS}

\vspace{10pt}

\begin{enumerate}
\item
The Feynman diagrams for the decays
$\tau^+\rta \ell^+\ell^+\ell^-\nu_\ell\overline \nu_\tau$ where the
$\ell$ are electrons or muons.

\item (a) The distribution in $z_{12} = \hat{p}_1\cdot \hat{p}_2$
for the two-like sign electrons (dashed line) and
in $z_{13}=\hat{p}_1\cdot\hat{p}_3$ for
the two-opposite sign leptons (solid line).  For comparison the corresponding
distribution for the phase space is also plotted (dash-dot line).

(b) The distributions in the total energy
$E_c=(E_{1}+E_{2}+E_{3})/m_\tau$ (solid line), and the
corresponding distribution for the phase space (dashed line).

\item
The branching ratio, $R$, for $\mu\rta e+\gamma +\nu+\overline{\nu},~\tau\rta
e+\gamma
+\nu+\overline{\nu}$ and $\tau\rta\mu+\gamma +\nu+\overline{v}$ as a function
of the
minimum detected photon energy, $\om_0$.  $y_0=\om_0/\om_{{\rm  Max}}$.  Note
that $R$ is
defined relative to the three body process $\mu\rta e+\nu+\overline{\nu},
\tau\rta
e+\nu+\overline\nu$ or
$\tau\rta\mu+\nu+\overline{\nu}$ and thus, for the $\tau$ decays, is not the
absolute
branching ratio.

\end{enumerate}

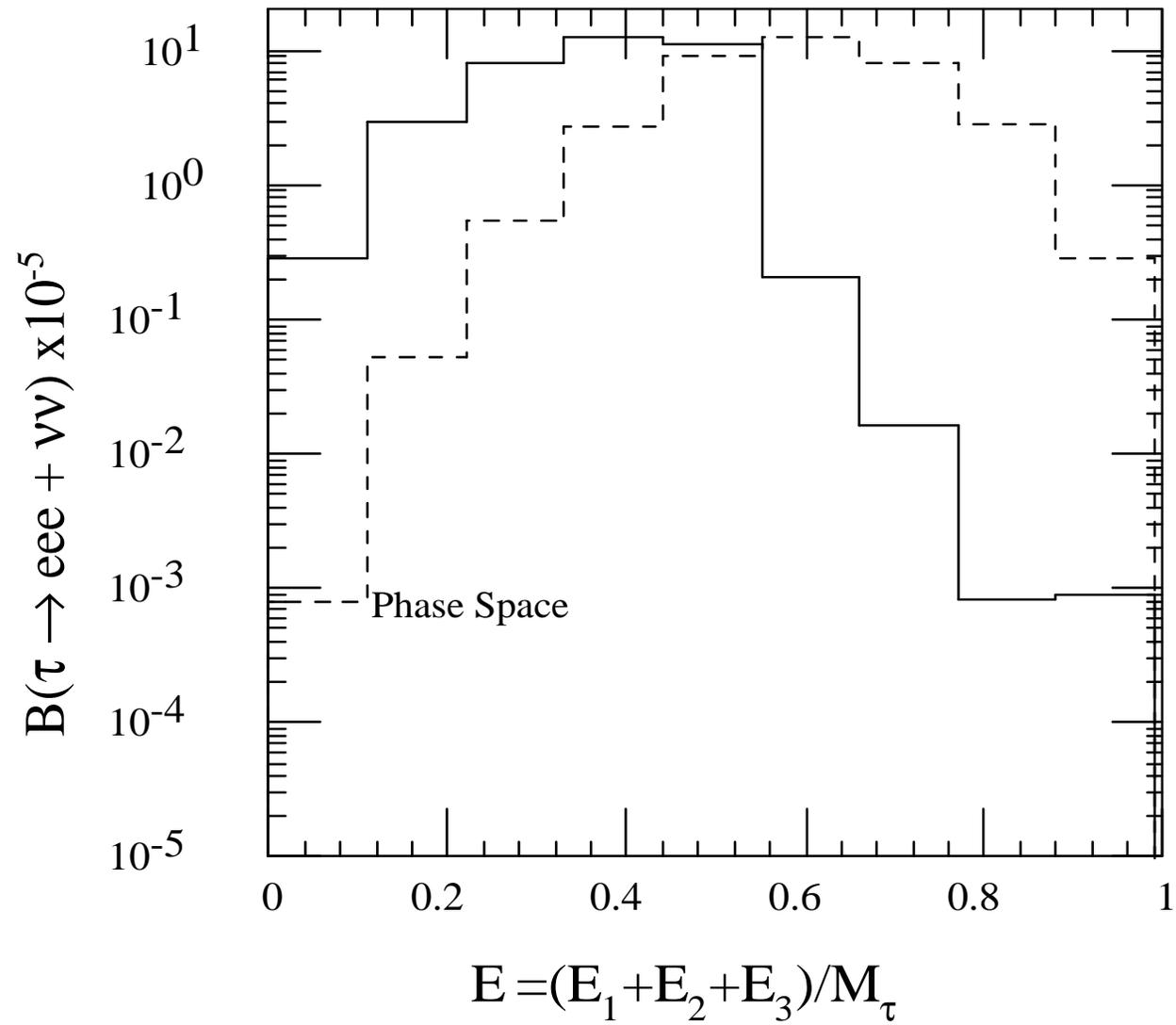
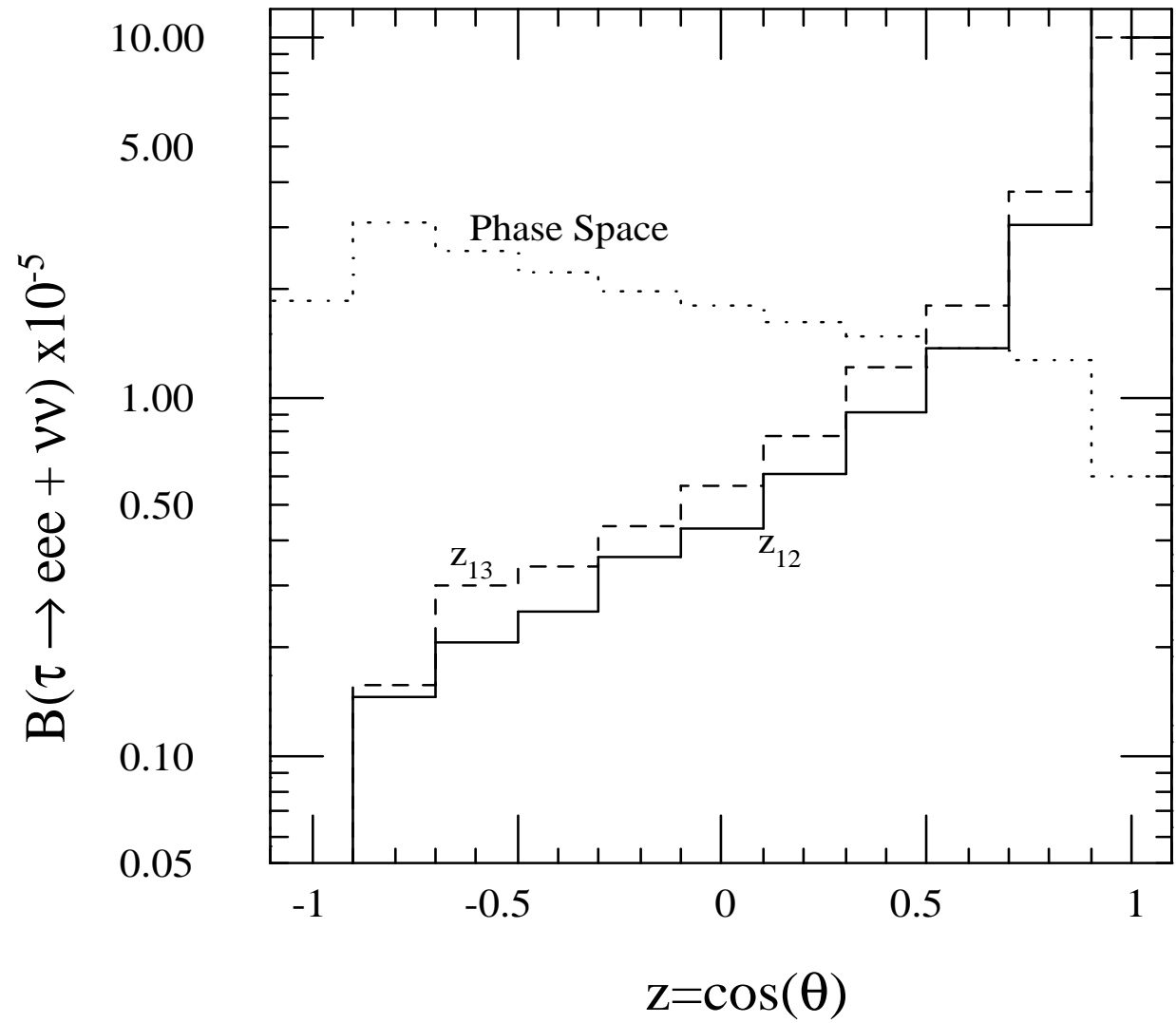
\end{document}